\documentclass[english,aip,preprint,nofootinbib,showpacs]{revtex4}
\usepackage[T1]{fontenc}
\usepackage[latin9]{inputenc}
\usepackage{float}
\usepackage{amsmath}
\usepackage{amssymb}
\usepackage{esint}

\makeatletter

\providecommand{\tabularnewline}{\\}

\@ifundefined{textcolor}{}
{%
 \definecolor{BLACK}{gray}{0}
 \definecolor{WHITE}{gray}{1}
 \definecolor{RED}{rgb}{1,0,0}
 \definecolor{GREEN}{rgb}{0,1,0}
 \definecolor{BLUE}{rgb}{0,0,1}
 \definecolor{CYAN}{cmyk}{1,0,0,0}
 \definecolor{MAGENTA}{cmyk}{0,1,0,0}
 \definecolor{YELLOW}{cmyk}{0,0,1,0}
}

\makeatother

\usepackage{babel}
\begin{document}

\title{Solutions associated with the point symmetries of the hyperbolic
Ernst equation}

\author{Sebastian Moeckel}

\affiliation{Theoretisch Physikalisches Institut, Friedrich-Schiller-Universität
Jena, Max-Wien-Platz 1, 07743 Jena, Germany}

\email{sebastian.moeckel@uni-jena.de}

\begin{abstract}
\noindent The continuous point symmetry algebra of the hyperbolic
Ernst equation is presented. In a second step the corresponding group
transformations are considered. Accordingly, the solutions of the
hyperbolic Ernst equation that are invariant under Lie point symmetries,
are constructed from the related invariant surface conditions. Furthermore,
all these solutions are revealed to be related to solutions of the
Euler-Poisson-Darboux equation by a simple coordinate transformation.
The parallels of these results to coordinate transformations, which
are important in the context of colliding plane wave space times,
are pointed out.
\end{abstract}

\pacs{02.20.Sv, 02.20.Tw, 02.30.Jr, 04.30.-w}

\maketitle

\section{Introduction}

Symmetry based methods can be easily employed for finding explicit
solutions of PDEs, especially if no supplementary requirements (e.g.
boundary conditions) are posed on the desired solution%
\footnote{The method can be also extended to boundary value problems, but here
also admitted symmetries of the boundary conditions have to be taken
into account.%
}. The Lie symmetries of a given set of PDEs can be used for locally
reducing the number of independent coordinates and therefore finding
explicit special solutions of the underlying set (an introduction
can be found in \cite{bluman2002symmetry}). This feature appears
to be of particular interest for a pair of PDEs depending on only
two independent coordinates. Here, a local reduction of independent
coordinates reduces the number of independent variables to one - hence
one is left with a system of ODEs.\\
In this article, the Lie point symmetry algebra of the hyperbolic
Ernst equation is presented. Furthermore, the solutions related to
these Lie generators are derived. Finally, some special remarks on
the relation of the so found solutions to the point symmetries of
the metric tensor for colliding gravitational plane waves are given.

\section{The hyperbolic Ernst equation and its point symmetries}

The hyperbolic Ernst equation can be written in the following form
\begin{eqnarray}
\left(Z+\bar{Z}\right)\left[2Z_{fg}+\frac{Z_{f}+Z_{g}}{f+g}\right] & = & 4Z_{f}Z_{g}\label{eq:f_g_hyp_ernst}
\end{eqnarray}
where $Z:\;\mathbb{R}\times\mathbb{R}\rightarrow\mathbb{C}$ is a
complex function of the two independent coordinates.\\
For deriving the Lie point symmetries, it is convenient to separate
(\ref{eq:f_g_hyp_ernst}) into a pair of real valued PDEs:
\begin{eqnarray}
K\left[2K_{fg}+\frac{K_{f}+K_{g}}{f+g}\right] & = & 2\left(K_{f}K_{g}-L_{f}L_{g}\right),\label{eq:Ernst_K}\\
K\left[2L_{fg}+\frac{L_{f}+L_{g}}{f+g}\right] & = & 2\left(K_{f}L_{g}+K_{g}L_{f}\right).\label{eq:Ernst_L}
\end{eqnarray}
where $Z=K+iL$.\\
The Lie point symmetry generators of the system (\ref{eq:Ernst_K})-(\ref{eq:Ernst_L})
can be determined rather easily by using an appropriate computer algebra
system (the package SYM \cite{dimas2004sym} has been used here):
\begin{alignat}{3}
X_{1} & =\partial_{f}-\partial_{g}, & \quad & X_{2} &  & =f\partial_{f}+g\partial_{g},\label{eq:Ernst_X1_X2}\\
X_{3} & =\partial_{L}, & \quad & X_{4} &  & =K\partial_{K}+L\partial_{L},\label{eq:Ernst_X3_X4}\\
X_{5} & =2KL\partial_{K}+\left(L^{2}-K^{2}\right)\partial_{L},\label{eq:Ernst_X5}
\end{alignat}
with the regarding commutator table\\
\begin{table}[H]
\noindent \begin{centering}
\begin{tabular}{c|c|c|c|c|c|}
$\left[\cdot,\cdot\right]$ & $X_{1}$ & $X_{2}$ & $X_{3}$ & $X_{4}$ & $X_{5}$\tabularnewline
\hline 
$X_{1}$ & $0$ & $X_{1}$ & \multicolumn{1}{c}{} & \multicolumn{1}{c}{} & \tabularnewline
\cline{1-3} 
$X_{2}$ & $-X_{1}$ & $0$ & \multicolumn{1}{c}{} & \multicolumn{1}{c}{} & \tabularnewline
\hline 
$X_{3}$ & \multicolumn{1}{c}{} &  & $0$ & $X_{3}$ & $2X_{4}$\tabularnewline
\cline{1-1} \cline{4-6} 
$X_{4}$ & \multicolumn{1}{c}{} &  & $-X_{3}$ & $0$ & $X_{5}$\tabularnewline
\cline{1-1} \cline{4-6} 
$X_{5}$ & \multicolumn{1}{c}{} &  & $-2X_{4}$ & $-X_{5}$ & $0$\tabularnewline
\hline 
\end{tabular}
\par\end{centering}

\caption{commutator table of the Lie algebra (\ref{eq:Ernst_X1_X2})-(\ref{eq:Ernst_X5})}
\end{table}
Hence, the Lie algebra decomposes into a two-dimensional non-abelian
sub-algebra and another three-dimensional sub-algebra. The Lie group
related to the two dimensional sub-algebra can be immediately classified
as being isomorphic to $Aff\left(1\right)$%
\footnote{The group of invertible affine transformations from $\mathbb{R}$
to $\mathbb{R}$.%
}, since there is only one real non-abelian two-dimensional Lie group
up to isomorphisms. The group corresponding to the three-dimensional
sub-algebra will be classified by considering its group action.\\
The corresponding group actions are obtained as
\begin{eqnarray}
\left(\begin{array}{c}
K\left(f,g\right)\\
L\left(f,g\right)
\end{array}\right) & \overset{e^{\beta X_{1}}}{\longrightarrow} & \left(\begin{array}{c}
K\left(f+\beta,g-\beta\right)\\
L\left(f+\beta,g-\beta\right)
\end{array}\right),\label{eq:K_L_trans}\\
\left(\begin{array}{c}
K\left(f,g\right)\\
L\left(f,g\right)
\end{array}\right) & \overset{e^{\alpha X_{2}}}{\longrightarrow} & \left(\begin{array}{c}
K\left(e^{\alpha}f,e^{\alpha}g\right)\\
L\left(e^{\alpha}f,e^{\alpha}g\right)
\end{array}\right),\label{eq:K_L_shift_arg}\\
\left(\begin{array}{c}
K\left(f,g\right)\\
L\left(f,g\right)
\end{array}\right) & \overset{e^{\gamma X_{3}}}{\longrightarrow} & \left(\begin{array}{c}
K\left(f,g\right)\\
L\left(f,g\right)+\gamma
\end{array}\right),\label{eq:L_shift}\\
\left(\begin{array}{c}
K\left(f,g\right)\\
L\left(f,g\right)
\end{array}\right) & \overset{e^{\delta X_{4}}}{\longrightarrow} & \left(\begin{array}{c}
e^{\delta}K\left(f,g\right)\\
e^{\delta}L\left(f,g\right)
\end{array}\right),\label{eq:K_L_scale}
\end{eqnarray}
where $\alpha$, $\beta$, $\gamma$, $\delta\in\mathbb{R}$. The
group action of $X_{5}$ is a bit more involved
\begin{eqnarray}
K & \overset{e^{\epsilon X_{5}}}{\longrightarrow} & \frac{K}{1-2L\epsilon+\left(K^{2}+L^{2}\right)\epsilon^{2}},\label{eq:X_5action_K}\\
L & \overset{e^{\epsilon X_{5}}}{\longrightarrow} & \frac{L-\left(K^{2}+L^{2}\right)\epsilon}{1-2L\epsilon+\left(K^{2}+L^{2}\right)\epsilon^{2}}.\label{eq:X_5action_l}
\end{eqnarray}
Combining the group actions of $X_{1}$ and $X_{2}$ on $Z=K+iL$
gives
\begin{alignat}{3}
Z\left(f,g\right) & \overset{e^{\alpha X_{2}}}{\longrightarrow} & \; Z\left(e^{\alpha}f,e^{\alpha}g\right) & \overset{e^{\beta X_{1}}}{\longrightarrow} & \; Z\left(e^{\alpha}f+\beta,e^{\alpha}g-\beta\right),\label{eq:Z_action_12}
\end{alignat}
where $\alpha,$ $\beta\in\mathbb{R}$.\\
The action of $X_{5}$ maps a given solution $Z$ of the Ernst equation
to a new solution according to
\begin{eqnarray}
Z & \overset{e^{\epsilon X_{5}}}{\longrightarrow} & \frac{Z-iZ\bar{Z}\epsilon}{1+i\left(Z-\bar{Z}\right)\epsilon+Z\bar{Z}\epsilon^{2}}=\frac{Z}{1+i\epsilon Z},\label{eq:Z_action_5}
\end{eqnarray}
where $\epsilon\in\mathbb{R}$. Finally, combining the actions of
$X_{3}$, $X_{4}$, and $X_{5}$ results in
\begin{alignat}{4}
Z & \overset{e^{\delta X_{4}}}{\longrightarrow} & \; e^{\delta}Z & \overset{e^{\gamma X_{3}}}{\longrightarrow} & \; e^{\delta}Z+i\gamma & \overset{e^{\epsilon X_{5}}}{\longrightarrow} & \;\frac{e^{\delta}Z+i\gamma}{1-\epsilon\gamma+i\epsilon e^{\delta}Z}=\; i\frac{aZ+ib}{cZ+id},\label{eq:X3_X4_X5_comb}
\end{alignat}
where 
\begin{alignat}{4}
a & = & e^{\delta/2}, & \quad & b & = &  & \gamma e^{-\delta/2},\label{eq:ab}\\
c & = & -\epsilon e^{\delta/2}, & \quad & d & = &  & \left(1-\epsilon\gamma\right)e^{-\delta/2}.\label{eq:cd}
\end{alignat}
Hence, $ad-bc=1$ and therefore the Lie group corresponding to $X_{3}$,
$X_{4}$, and $X_{5}$ is isomorphic to $SL\left(2,\mathbb{R}\right)$.
As a special remark, the transformation (\ref{eq:X3_X4_X5_comb})
has already been considered by Neugebauer and Kramer in 1969 (\cite{neugebauer1969methode}).
In the context of space-times with two commuting global Killing vector
fields they have shown, that the transformation (\ref{eq:X3_X4_X5_comb})
can be derived by a simple $SL\left(2,\mathbb{R}\right)$ rotation
of the two Killing vector fields.\\
Hence, the Ernst equation (\ref{eq:Ernst_K})-(\ref{eq:Ernst_L})
is invariant under a Lie group, whose algebra is isomorphic to $aff\left(1\right)\oplus sl\left(2,\mathbb{R}\right)$.

\section{Special solutions associated to the point symmetries}

\subsection{Solutions related to $X_{1}=\partial_{g}-\partial_{f}$}

The invariant surface conditions for the solutions $K$ and $L$ of
the system of PDEs (\ref{eq:Ernst_K})-(\ref{eq:Ernst_L}) with respect
to the infinitesimal generator $X_{1}=\partial_{g}-\partial_{f}$
read
\begin{eqnarray}
\frac{\partial K}{\partial g} & = & \frac{\partial K}{\partial f},\label{eq:inv_surface_K1}\\
\frac{\partial L}{\partial g} & = & \frac{\partial L}{\partial f}.\label{eq:inv_surface_L1}
\end{eqnarray}
Plugging the constraints (\ref{eq:inv_surface_K1})-(\ref{eq:inv_surface_L1})
into the system of PDEs (\ref{eq:Ernst_K})-(\ref{eq:Ernst_L}) yields
the following system of (parametric) ordinary differential equations
in $K$ and $L$ with the independent coordinate $f$ and the parameter
$g$ 
\begin{eqnarray}
K\left[K_{ff}+\frac{K_{f}}{f+g}\right] & = & K_{f}^{2}-L_{f}^{2},\label{eq:inv_ode1}\\
K\left[L_{ff}+\frac{L_{f}}{f+g}\right] & = & 2K_{f}L_{f}.\label{eq:inv_ode2}
\end{eqnarray}
The general solution of (\ref{eq:inv_ode1})-(\ref{eq:inv_ode2})
can be found as follows.\\
First, it is necessary to write (\ref{eq:inv_ode1}) in solved form
\begin{eqnarray}
K_{ff} & = & \frac{K_{f}^{2}-L_{f}^{2}}{K}-\frac{K_{f}}{f+g}.\label{eq:K_ff}
\end{eqnarray}
Furthermore, solving (\ref{eq:inv_ode1}) for $L_{f}^{2}$ yields
\begin{eqnarray}
L_{f}^{2} & = & K_{f}^{2}-K\left(K_{ff}+\frac{K_{f}}{f+g}\right),\label{eq:L_f^2}
\end{eqnarray}
and solving (\ref{eq:inv_ode2}) for $L_{ff}$ gives
\begin{eqnarray}
L_{ff} & = & L_{f}\left(2\frac{K_{f}}{K}-\frac{1}{f+g}\right).\label{eq:L_ff}
\end{eqnarray}
Differentiating (\ref{eq:K_ff}) with respect to $f$ leads to
\begin{eqnarray}
K_{fff} & = & -K_{f}\left(\frac{K_{f}^{2}-L_{f}^{2}}{K^{2}}\right)+2\left(\frac{K_{f}K_{ff}-L_{f}L_{ff}}{K}\right)-\frac{K_{ff}}{f+g}+\frac{K_{f}}{\left(f+g\right)^{2}}.\label{eq:K_fff}
\end{eqnarray}
Now replacing $L_{ff}$ and $L_{f}^{2}$ according to (\ref{eq:L_ff})
and (\ref{eq:L_f^2}) yields the following third order nonlinear ordinary
differential equation in $K$
\begin{alignat}{3}
K_{fff} & = & F\left(f,K,K_{f},K_{ff}\right) & = & -4\frac{K_{f}^{3}}{K^{2}}-3\frac{K_{ff}}{f+g}-\frac{K_{f}}{\left(f+g\right)^{2}}+5\frac{K_{f}K_{ff}}{K}+5\frac{K_{f}^{2}}{K\left(f+g\right)}.\label{eq:K_fff_ode}
\end{alignat}
Note that $g$ appears only as a parameter in (\ref{eq:K_fff_ode}).
The method of determining integrating factors for generating corresponding
first integrals (see the appendix and \cite{bluman2002symmetry} for
details) is one possible way for obtaining the general solution of
(\ref{eq:K_fff_ode}). For this purpose, the surface generated by
equation (\ref{eq:K_fff_ode}) in the 2+3 dimensional jet-space%
\footnote{The jet-space is constructed by interpreting the independent variable
$f$ and the dependent variable $K$ together with all derivatives
appearing in the ODE under consideration (i.e. $K_{f}$, $K_{ff}$,
and $K_{fff}$) as independent coordinates of a 2+3 dimensional space.
See \cite{bluman2002symmetry} for details. %
} with coordinates $\left(f,K,K_{1},K_{2},K_{3}\right)$ is considered
\begin{alignat}{3}
K_{3} & = & F\left(f,K,K_{1},K_{2}\right) & = & -4\frac{K_{1}^{3}}{K^{2}}-3\frac{K_{2}}{f+g}-\frac{K_{1}}{\left(f+g\right)^{2}}+5\frac{K_{1}K_{2}}{K}+5\frac{K_{1}^{2}}{K\left(f+g\right)}.\label{eq:K_fff_ode-1-1}
\end{alignat}
Since (\ref{eq:K_fff_ode-1-1}) does not involve higher order terms
of $K_{2}$, the following ansatz for the integrating factor seems
to be promising
\begin{eqnarray}
\Lambda\left(f,K,K_{1}\right) & = & \alpha\left(f,K\right)+K_{1}\beta\left(f,K\right).\label{eq:int_factor}
\end{eqnarray}
Plugging (\ref{eq:int_factor}) into the integrating-factor determining
equation (\ref{eq:3rd_ODE_int1}) immediately yields $\beta\left(f,K\right)\equiv0$.
Plugging the remaining expression for $\Lambda$ into the second determining
equation (\ref{eq:3rd_ODE_int2}) yields an overdetermined system
of six linear PDEs for $\alpha$
\begin{eqnarray}
\frac{9}{K^{2}}\alpha+\frac{15}{K}\alpha_{K}+3\alpha_{KK} & = & 0,\label{eq:a_1}\\
-\frac{6}{K^{3}}\alpha-\frac{2}{K^{2}}\alpha_{K}+\frac{5}{K}\alpha_{KK}+\alpha_{KKK} & = & 0,\label{eq:a_2}\\
-\frac{10}{\left(f+g\right)K}\alpha-\frac{6}{f+g}\alpha_{K}+\frac{5}{K}\alpha_{f}+3\alpha_{fK} & = & 0,\label{eq:a_3}\\
-\frac{8}{\left(f+g\right)^{3}}\alpha+\frac{7}{\left(f+g\right)^{2}}\alpha_{f}-\frac{3}{f+g}\alpha_{ff}+\alpha_{fff} & = & 0,\label{eq:a_4}\\
\left.\begin{array}{c}
\frac{5}{\left(f+g\right)K^{2}}\alpha-\frac{5}{\left(f+g\right)K}\alpha_{K}-\frac{3}{f+g}\alpha_{KK}+\frac{2}{K^{2}}\alpha_{f}\\
+\frac{10}{K}\alpha_{fK}+3\alpha_{fKK}
\end{array}\right\}  & = & 0,\label{eq:a_5}\\
\left.\begin{array}{c}
\frac{10}{\left(f+g\right)^{2}K}\alpha+\frac{6}{\left(f+g\right)^{2}}\alpha_{K}-\frac{10}{\left(f+g\right)K}\alpha_{f}-\frac{6}{\left(f+g\right)}\alpha_{fK}\\
+\frac{5}{K}\alpha_{ff}+3\alpha_{ffK}
\end{array}\right\}  & = & 0.\label{eq:a_6}
\end{eqnarray}
Solving (\ref{eq:a_1}) yields
\begin{eqnarray}
\alpha\left(f,K\right) & = & \frac{C_{1}\left(f\right)}{K}+\frac{C_{2}\left(f\right)}{K^{3}},\label{eq:a_1_sol}
\end{eqnarray}
where $C_{1}\left(f\right)$ and $C_{2}\left(f\right)$ are arbitrary
functions of $f$. Furthermore, the expression (\ref{eq:a_1_sol})
solves (\ref{eq:a_2}) identically. Plugging (\ref{eq:a_1_sol}) into
(\ref{eq:a_3}) yields the following determining equations for $C_{1}\left(f\right)$
and $C_{2}\left(f\right)$
\begin{eqnarray}
2C_{1}\left(f\right)-\left(f+g\right)C_{1}^{\prime}\left(f\right) & = & 0,\label{eq:C1}\\
2C_{2}\left(f\right)-\left(f+g\right)C_{2}^{\prime}\left(f\right) & = & 0,\label{eq:C2}
\end{eqnarray}
which yield the solutions
\begin{eqnarray}
C_{1}\left(f\right) & = & c_{1}\left(f+g\right)^{2},\label{eq:C1_sol}\\
C_{2}\left(f\right) & = & c_{2}\left(f+g\right)^{2},\label{eq:C2_sol}
\end{eqnarray}
where $c_{1},c_{2}\in\mathbb{C}$ for fixed $g$. Hence $\alpha$
reads
\begin{eqnarray}
\alpha\left(f,K\right) & = & c_{1}\frac{\left(f+g\right)^{2}}{K}+c_{2}\frac{\left(f+g\right)^{2}}{K^{3}}.\label{eq:a_1_sol-1}
\end{eqnarray}
Expression (\ref{eq:a_1_sol-1}) also solves (\ref{eq:a_4})-(\ref{eq:a_6})
identically. Hence, two functionally independent integrating factors
for equation (\ref{eq:K_fff_ode}) are obtained with
\begin{alignat}{4}
\Lambda_{1}\left(f,K\right) & = & \frac{\left(f+g\right)^{2}}{K}, & \quad & \Lambda_{2}\left(f,K\right) & = & \frac{\left(f+g\right)^{2}}{K^{3}}.\label{eq:int_factor_sol}
\end{alignat}
Now plugging $\Lambda_{1}$ and $\Lambda_{2}$ into the line integral
formula (\ref{eq:first_int_path}) for determining the corresponding
first integrals give
\begin{eqnarray}
\psi_{1}\left(f,K,K_{1},K_{2}\right) & = & \frac{\left(f+g\right)}{K^{2}}\left\{ K_{1}\left[K-2\left(f+g\right)K_{1}\right]+\left(f+g\right)KK_{2}\right\} ,\label{eq:psi1}\\
\psi_{2}\left(f,K,K_{1},K_{2}\right) & = & \frac{\left(f+g\right)}{K^{4}}\left\{ K_{1}\left[K-\left(f+g\right)K_{1}\right]+\left(f+g\right)KK_{2}\right\} ,\label{eq:psi2}
\end{eqnarray}
where the path of integration has been chosen parallel to the axis
of $\mathbb{R}^{4}$ from $\left(0,\tilde{K},0,0\right)$ to $\left(f,K,K_{1},K_{2}\right)$,
such that the singularity at $K=0$ is avoided. Since (\ref{eq:psi1})
and (\ref{eq:psi2}) are first integrals of (\ref{eq:K_fff_ode}),
every solution of (\ref{eq:K_fff_ode}) satisfies $\psi_{1}\left(f,K,K_{f},K_{ff}\right)=c_{1}=\textrm{const.}$,
together with $\psi_{2}\left(f,K,K_{f},K_{ff}\right)=c_{2}=\textrm{const.}$
for arbitrary constants $c_{1},c_{2}\in\mathbb{C}$ (for fixed $g$).
Therefore, setting $\psi_{1}=c_{1}$ and $\psi_{2}=c_{2}$ results
in an effective reduction of the third order equation (\ref{eq:K_fff_ode})
to the first order equation
\begin{eqnarray}
\frac{c_{1}K^{2}+\left(f+g\right)^{2}K_{f}^{2}}{K^{4}} & = & c_{2}.\label{eq:1st_ode}
\end{eqnarray}
Because (\ref{eq:1st_ode}) is separable, a general solution can be
obtained by quadrature, leading to the general solutions
\begin{eqnarray}
K_{I}\left(f,g\right) & =\pm & \sqrt{\frac{c_{1}\left(g\right)}{c_{2}\left(g\right)}}\csc\left[\sqrt{c_{1}\left(g\right)}\left(c_{3}\left(g\right)-\ln\left(f+g\right)\right)\right],\label{eq:sol_K_1}\\
K_{II}\left(f,g\right) & =\pm & \sqrt{\frac{c_{1}\left(g\right)}{c_{2}\left(g\right)}}\csc\left[\sqrt{c_{1}\left(g\right)}\left(c_{3}\left(g\right)+\ln\left(f+g\right)\right)\right],\label{eq:sol_K_2}
\end{eqnarray}
where $c_{1}$, $c_{2}$ and $c_{3}$ are still arbitrary functions
of $g$ (recall that ODE (\ref{eq:K_fff_ode}) is parametric in $g$).
Inserting (\ref{eq:sol_K_1}) and (\ref{eq:sol_K_2}) into (\ref{eq:L_f^2})
yields the following expressions for $L\left(f,g\right)$
\begin{eqnarray}
L_{I}\left(f,g\right) & = & \pm\sqrt{-\frac{c_{1}\left(g\right)}{c_{2}\left(g\right)}}\cot\left[\sqrt{c_{1}\left(g\right)}\left(c_{3}\left(g\right)-\ln\left(f+g\right)\right)\right]+c_{4}\left(g\right),\label{eq:sol_L_1}\\
L_{II}\left(f,g\right) & = & \pm\sqrt{-\frac{c_{1}\left(g\right)}{c_{2}\left(g\right)}}\cot\left[\sqrt{c_{1}\left(g\right)}\left(c_{3}\left(g\right)+\ln\left(f+g\right)\right)\right]+c_{4}\left(g\right).\label{eq:sol_L_2}
\end{eqnarray}
Plugging the expressions for $L$ and $K$ back into the invariant
surface conditions (\ref{eq:inv_surface_K1})-(\ref{eq:inv_surface_L1})
allows for the determination of the remaining functions $c_{1}\left(g\right)$,
$c_{2}\left(g\right)$, $c_{3}\left(g\right)$ and $c_{4}\left(g\right)$.
As a result, all functions $c_{1}\left(g\right)$, $c_{2}\left(g\right)$,
$c_{3}\left(g\right)$ and $c_{4}\left(g\right)$ need to be simple
constants, thus generating the following solutions
\begin{eqnarray}
K_{I}\left(f,g\right) & =\pm & \sqrt{\frac{c_{1}}{c_{2}}}\csc\left[\sqrt{c_{1}}\left(c_{3}-\ln\left(f+g\right)\right)\right],\label{eq:sol_K_1-1}\\
K_{II}\left(f,g\right) & =\pm & \sqrt{\frac{c_{1}}{c_{2}}}\csc\left[\sqrt{c_{1}}\left(c_{3}+\ln\left(f+g\right)\right)\right],\label{eq:sol_K_2-1}
\end{eqnarray}
\begin{eqnarray}
L_{I}\left(f,g\right) & = & \pm\sqrt{-\frac{c_{1}}{c_{2}}}\cot\left[\sqrt{c_{1}}\left(c_{3}-\ln\left(f+g\right)\right)\right]+c_{4},\label{eq:sol_L_1-1}\\
L_{II}\left(f,g\right) & = & \pm\sqrt{-\frac{c_{1}}{c_{2}}}\cot\left[\sqrt{c_{1}}\left(c_{3}+\ln\left(f+g\right)\right)\right]+c_{4},\label{eq:sol_L_2-1}
\end{eqnarray}
where $c_{1}$, $c_{2}$, $c_{3}$, $c_{4}\in\mathbb{C}$.\\
For interpreting $K$ and $L$ as the real- and imaginary parts of
the Ernst potential $Z=K+iL$, it is necessary to transform the expressions
(\ref{eq:sol_K_1-1})-(\ref{eq:sol_L_2-1}) to real valued functions.
One possibility for achieving this goal is to choose $c_{3}=\frac{1}{\sqrt{c_{1}}}\frac{\pi}{2}$,
$\sqrt{c_{1}}=iA$, $\sqrt{c_{2}}=iB$, $c_{4}=C$ with $A,B,C\in\mathbb{R}$,
resulting in
\begin{eqnarray}
K\left(f,g\right) & = & \frac{2A}{B}\frac{\left(f+g\right)^{A}}{1+\left(f+g\right)^{2A}},\label{eq:sol_K_A_B_C}\\
L\left(f,g\right) & = & \frac{A}{B}\frac{1-\left(f+g\right)^{2A}}{1+\left(f+g\right)^{2A}}+C,\label{eq:sol_L_A_B_C}
\end{eqnarray}
where a possible factor of $-1$ has been absorbed in the constants.

\subsection{Solutions related to $X_{2}=f\partial_{f}+g\partial_{g}$}

The invariant surface conditions for the solutions $K$ and $L$ of
the system of PDEs (\ref{eq:Ernst_K})-(\ref{eq:Ernst_L}) with respect
to the infinitesimal generator $X_{2}=f\partial_{f}+g\partial_{g}$
read
\begin{eqnarray}
g\frac{\partial K}{\partial g} & = & -f\frac{\partial K}{\partial f},\label{eq:inv_surface_K1-1}\\
g\frac{\partial L}{\partial g} & = & -f\frac{\partial L}{\partial f}.\label{eq:inv_surface_L1-1}
\end{eqnarray}
Now, following exactly the same steps as for the solutions corresponding
to $X_{1}$, leads to the final solution
\begin{eqnarray}
K\left(f,g\right) & = & \frac{A}{B}\textrm{sech}\left[2A\arctan\left(\sqrt{\frac{f}{g}}\right)\right],\label{eq:K_L2}\\
L\left(f,g\right) & = & \frac{A}{B}\tanh\left[2A\arctan\left(\sqrt{\frac{f}{g}}\right)\right]+C,\label{eq:L_L2}
\end{eqnarray}
where $A$, $B$, $C\in\mathbb{R}$.

\section{Relation to gravitational plane wave collisions}

The two group invariant solutions obtained in the preceding section
appear to be surprisingly similar, since they are both of the form
\begin{eqnarray}
K\left(f,g\right) & = & \textrm{sech}\left[F\left(f,g\right)\right],\label{eq:allg_K}\\
L\left(f,g\right) & = & \tanh\left[F\left(f,g\right)\right],\label{eq:allg_L}
\end{eqnarray}
where $F\left(f,g\right)$ is an arbitrary real function of $f$ and
$g$ and all pre-factors have been omitted, which can be re-introduced
later on by using the apparent scaling symmetry of the Ernst equation.\\
Plugging the ansatz (\ref{eq:allg_K})-(\ref{eq:allg_L}) into the
separated form of the Ernst equation (\ref{eq:Ernst_K}) and (\ref{eq:Ernst_L})
leads to the following result
\begin{eqnarray}
\frac{\textrm{sech}\left[F\right]\tanh\left[F\right]}{f+g}\left\{ 2\left(f+g\right)F_{fg}+F_{f}+F_{g}\right\}  & = & 0,\label{eq:ernst_K_F}\\
\frac{\textrm{sech}\left[F\right]}{f+g}\left\{ 2\left(f+g\right)F_{fg}+F_{f}+F_{g}\right\}  & = & 0,\label{eq:ernst_L_F}
\end{eqnarray}
Hence, if $F\left(f,g\right)$ satisfies the EPD equation 
\begin{eqnarray}
2\left(f+g\right)F_{fg}+F_{f}+F_{g} & = & 0,\label{eq:EPD_F}
\end{eqnarray}
$K\left(f,g\right)$ and $L\left(f,g\right)$ solve the Ernst equations
(\ref{eq:Ernst_K})-(\ref{eq:Ernst_L}).\\
Thus, every real function $F\left(f,g\right)$ satisfying the EPD
equation (\ref{eq:EPD_F}) generates a complex valued Ernst potential
by means of 
\begin{eqnarray}
Z\left(f,g\right) & = & \textrm{sech}\left[F\left(f,g\right)\right]+i\tanh\left[F\left(f,g\right)\right],\label{eq:Z_F}
\end{eqnarray}
which is a solution of the Ernst equation.\\
$F$ can be interpreted as the scalar potential of a real valued Ernst
potential $Z_{o}=\exp\left(2F\right)$. Therefore, $Z_{o}$ is possibly
the potential of a collinear colliding gravitational plane wave solution,
if certain conditions are satisfied (see \cite{griffiths1991colliding}
for a detailed discussion of colliding gravitational plane wave space-times).-\\
Writing (\ref{eq:Z_F}) in terms of $2F=\ln Z_{o}$ leads to the following
form of the transformation
\begin{eqnarray}
Z & = & \frac{1+iZ_{o}}{i+Z_{o}},\label{eq:Z_from_Zo}
\end{eqnarray}
revealing (\ref{eq:Z_F}) to be just a special case of the transformation
induced by the Lie sub-algebra generated by $X_{3}$, $X_{4}$ and
$X_{5}$ (c.f. (\ref{eq:X3_X4_X5_comb})).\\
As an additional feature, transformation (\ref{eq:X3_X4_X5_comb})
leaves the plane-wave conditions invariant (cf. \cite{griffiths1991colliding}).
Therefore colliding wave solutions are mapped to colliding wave solutions
by transformation (\ref{eq:Z_from_Zo}).\\
Now, recalling that (\ref{eq:X3_X4_X5_comb}) can be interpreted as
a simple rotation of the two globally commuting Killing vector fields%
\footnote{Colliding gravitational plane wave space-times are equipped with a
pair of space-like globally commuting Killing vector fields.%
}, this leads to the conclusion that the two group-invariant solutions
(\ref{eq:sol_K_A_B_C})-(\ref{eq:sol_L_A_B_C}) and (\ref{eq:K_L2})-(\ref{eq:L_L2}),
as well as all solutions generated by using (\ref{eq:X3_X4_X5_comb}),
cannot be considered as inherently non-collinear%
\footnote{Solutions that lead to a reduction of the Ernst equation to the linear
Euler-Poisson-Darboux equation are classified as collinear solutions
(see \cite{griffiths1991colliding} for more details).%
}, since they can be reduced to collinear solutions by a simple coordinate
transformation. It follows that the solutions of the Ernst equation,
which are invariant under continuous point transformations can all
be reduced to collinear solutions. This relation has already been
known for the point-symmetries generated by $X_{3}$, $X_{4}$ and
$X_{5}$ (cf. \cite{neugebauer1969methode}), but has not been obvious
for the solutions which have been generated from the invariant surface
condition with respect to $X_{1}$ and $X_{2}$.

\section{Conclusions and outlook}

As a major result, it has been shown that Lie's method for determining
symmetry invariant solutions, is well suited for generating explicit
solutions of the hyperbolic Ernst equation. Furthermore, the important
relation between the class of solutions generated from the closed
sub-algebra $\left\langle X_{1},X_{2}\right\rangle $ and solutions
that are obtained by only using transformations of the closed sub-algebra
$\left\langle X_{3},X_{4},X_{5}\right\rangle $ has been revealed.
Moreover, the relation to simple coordinate transformations in the
context of colliding plane wave space-times has been established.\\
A future perspective for generating new group-invariant solutions
is to consider contact- and higher order symmetries of the Ernst equation.
The resulting solutions can be obtained by invoking a similar procedure
as has been carried out for the point symmetries (cf. \cite{bluman2002symmetry}
and \cite{bluman2010applications}). In addition, the group-invariant
solutions of the coupled pair of hyperbolic Ernst equations, appearing
in the context of coupled electromagnetic and gravitational plane
wave collisions, can be considered.\\

\appendix

\section*{Appendix}

Taking the definition from \cite{bluman2002symmetry}, a first integral
of an $n$th-order ODE
\begin{eqnarray}
y^{\left(n\right)} & = & F\left(x,y,y^{\prime},\ldots,y^{\left(n-1\right)}\right)\label{eq:app_nth_order_ODE}
\end{eqnarray}
is a function $\psi\left(x,y,y^{\prime},\ldots,y^{\left(n-1\right)}\right)$
with an essential dependence on $y^{\left(n-1\right)}$ satisfying
\begin{eqnarray}
\frac{d\psi}{dx} & = & 0,\quad\textrm{when }\quad y^{\left(n\right)}=F,\label{eq:first_int}
\end{eqnarray}
i.e. $\psi\left(x,y,y^{\prime},\ldots,y^{\left(n-1\right)}\right)$
is constant for every solution $y=\Theta\left(x\right)$ of the ODE
(\ref{eq:app_nth_order_ODE}).\\
Hence, a first integral is a conserved quantity for each solution
and therefore the knowledge of $r$ functionally independent first
integrals leads to an effective reduction of the order of the given
ODE to $n-r$.\\
Furthermore, an integrating factor of ODE (\ref{eq:app_nth_order_ODE})
is defined as a function $\Lambda\left(x,y,y^{\prime},\ldots,y^{\left(l\right)}\right)\not\equiv0$
($0\leq l\leq n-1$, $l$ is the order of the integrating factor),
such that
\begin{eqnarray}
\Lambda\left(x,y,y^{\prime},\ldots,y^{\left(l\right)}\right)\left(y^{\left(n\right)}-F\left(x,y,y^{\prime},\ldots,y^{\left(n-1\right)}\right)\right) & = & \frac{d\psi}{dx}\left(x,y,y^{\prime},\ldots,y^{\left(n-1\right)}\right),\label{eq:app_int_factor}
\end{eqnarray}
for a function $\psi\left(x,y,y^{\prime},\ldots,y^{\left(n-1\right)}\right)$
with an essential dependence on $y^{\left(n-1\right)}$.\\
Considering a general third order ODE
\begin{eqnarray}
y^{\prime\prime\prime} & = & F\left(x,y,y^{\prime},y^{\prime\prime}\right),\label{eq:3rd_ODE}
\end{eqnarray}
represented by the surface
\begin{eqnarray}
y_{3} & = & F\left(x,y,y_{1},y_{2}\right).\label{eq:3rd_ODE_surf}
\end{eqnarray}
\cite{bluman2002symmetry} provides the explicit determining system
for integrating factors:
\begin{eqnarray}
2\Lambda_{y_{1}}+\Lambda_{y_{2}x}+y_{1}\Lambda_{y_{2}y}+y_{2}\Lambda_{y_{2}y_{1}}+\left(F\Lambda\right)_{y_{2}y_{2}} & = & 0,\label{eq:3rd_ODE_int1}\\
\left.\begin{array}{c}
3y_{2}\Lambda_{xy}+3y_{1}y_{2}\Lambda_{yy}+3y_{2}^{2}\Lambda_{yy_{1}}+\Lambda_{xxx}+y_{1}^{3}\Lambda_{yyy}+y_{2}^{3}\Lambda_{y_{1}y_{1}y_{1}}\\
+3y_{1}\Lambda_{xxy}+3y_{2}\Lambda_{xxy_{1}}+3y_{1}^{2}\Lambda_{xyy}+3y_{2}^{2}\Lambda_{xy_{1}y_{1}}+3y_{1}^{2}y_{2}\Lambda_{yyy_{1}}\\
+3y_{1}y_{2}^{2}\Lambda_{yy_{1}y_{1}}+6y_{1}y_{2}\Lambda_{xyy_{1}}+\left(F\Lambda\right)_{y}-\left(F\Lambda\right)_{xy_{1}}-y_{1}\left(F\Lambda\right)_{yy_{1}}\\
-y_{2}\left(F\Lambda\right)_{y_{1}y_{1}}+y_{2}\left(F\Lambda\right)_{yy_{2}}+\left(F\Lambda\right)_{xxy_{2}}+y_{1}^{2}\left(F\Lambda\right)_{yyy_{2}}\\
+y_{2}^{2}\left(F\Lambda\right)_{y_{1}y_{1}y_{2}}+2y_{1}\left(f\Lambda\right)_{xyy_{2}}+2y_{2}\left(f\Lambda\right)_{xy_{1}y_{2}}+2y_{1}y_{2}\left(F\Lambda\right)_{yy_{1}y_{2}}
\end{array}\right\}  & = & 0.\label{eq:3rd_ODE_int2}
\end{eqnarray}
For a general integrating factor $\Lambda\left(x,y,y_{1},y_{2}\right)$,
the system (\ref{eq:3rd_ODE_int1}) and (\ref{eq:3rd_ODE_int2}) admits
an infinite number of solutions, but restricting $\Lambda$ to special
ansatzes yields an overdetermined system of linear PDEs that has at
most a finite number of independent solutions. For instance, if $F\left(x,y,y_{1},y_{2}\right)$
is linear in $y_{2}$, the following ansatz leads to some remarkable
simplifications
\begin{eqnarray}
\Lambda\left(x,y,y_{1}\right) & = & \alpha\left(x,y\right)+y_{1}\beta\left(x,y\right).\label{eq:special_int}
\end{eqnarray}
Plugging (\ref{eq:special_int}) into (\ref{eq:3rd_ODE_int1}) immediately
yields $\beta\equiv0$, and (\ref{eq:3rd_ODE_int2}) decomposes into
several linear PDEs for determining $\alpha$.\\
When an integrating factor of (\ref{eq:3rd_ODE}) has been obtained,
a corresponding first integral can be calculated by means of the following
line integral formula (\cite{bluman2002symmetry})
\begin{eqnarray}
\psi & = & \intop_{C}\left\{ \left[-\left(F\Lambda\right)_{y_{1}}+\left(F\Lambda\right)_{xy_{2}}+y_{1}\left(F\Lambda\right)_{yy_{2}}+y_{2}\left(F\Lambda\right)_{y_{1}y_{2}}+y_{2}\Lambda_{y}\right.\right.\nonumber \\
 &  & \left.+\Lambda_{xx}+y_{1}^{2}\Lambda_{yy}+y_{2}^{2}\Lambda_{y_{1}y_{1}}+2y_{1}\Lambda_{xy}+2y_{2}\Lambda_{xy_{1}}+2y_{1}y_{2}\Lambda_{yy_{1}}\right]\left(dy-y_{1}dx\right)\nonumber \\
 &  & \left.-\left[\left(F\Lambda\right)_{y_{2}}+\Lambda_{x}+y_{1}\Lambda_{y}+y_{2}\Lambda_{y_{1}}\right]\left(dy_{1}-y_{2}dx\right)+\Lambda\left(dy_{2}-fdx\right)\right\} ,\label{eq:first_int_path}
\end{eqnarray}
where the path $C$ can be chosen arbitrarily from any point $\left(\tilde{x},\tilde{y},\tilde{y}_{1},\tilde{y}_{2}\right)$
to $\left(x,y,y_{1},y_{2}\right)$, such that all singularities are
avoided.

\bibliographystyle{aipnum4-1}
\bibliography{paper_lie}

\end{document}